# Learning to Exploit Different Translation Resources for Cross Language Information Retrieval

Hosein Azarbonyad
School of Electrical and Computer Engineering
College of Engineering
University of Tehran, Tehran, Iran
h.azarbonyad@ut.ac.ir

Azadeh Shakery
School of Electrical and Computer Engineering
College of Engineering
University of Tehran, Tehran, Iran
shakery@ut.ac.ir

Heshaam Faili
School of Electrical and Computer Engineering
College of Engineering
University of Tehran, Tehran, Iran
hfaili@ut.ac.ir

*Abstract*—One of the important factors that affects the performance of Cross Language Information Retrieval(CLIR) is the quality of translations being employed inCLIR. In order to improve the quality of translations, it is important toexploitavailable resources efficiently. Employing different translation resources with different characteristics has many challenges. In this paper, we propose a method for exploiting available translation resources simultaneously. This method employs Learning to Rank(LTR) for exploiting different translation resources. To apply LTR methods for query translation, we define different translation relation based features in addition to context based features. We use the contextual information contained in translation resources for extracting context based features.The proposed method uses LTR to construct a translation ranking model based on defined features. The constructed model is used for ranking translation candidates of query words. To evaluate the proposed method we do English-Persian CLIR, in which we employ the translation ranking model to find translations of English queries and employ the translations to retrieve Persian documents. Experimental results show that our approach significantly outperforms single resource based CLIR methods.

I. INTRODUCTION

With the advent of different web pages with different languages, World Wide Web turned into a multilingual environment. Considering the web as a multilingual environment, we need methods and search engines for retrieving documents in different languages. In fact, a query in a certain language could have different relevant documents in multiple languages. This challenge demands methods for retrieving documents in languages different form the language of query. Cross Language Information Retrieval (CLIR) deals with this problem. The main aim of CLIR is finding documents in a language different from the language of query. Since in CLIR the language of documents (target language) is different from query language (source language), we need translation methods for either translating documents to the language of queries (document translation) or translating queries to the language of documents (query translation).Also, we can translate documents and queries to an intermediate language and solve the problem of language difference [1,2]. Generally, document translation has better performance than query translation in CLIR. However, because of high cost and complexity of machine translation, document translation has not been widely used in CLIR [1, 2]. Therefore, this paper focuses on query translation approach.

Different translation methods have been used for translating queries in CLIR. Most of them use translation resources for extracting translations of query terms and constructing the query in target language. Translation resources used in CLIR could be categorized into four different categories [2]: 1) dictionaries, 2) comparable corpora, 3) parallel corpora, and 4) machine translators. As mentioned, using machine translators has high cost and complexity. Thus, in this research, we concentrate on dictionaries, comparable corpora, and parallel corpora for translating queries. Each of these resources has specifications that other resources may not have them. For example, parallel corpora usually are more accurate than comparable corpora and translations extracted from parallel corpora for query terms could be the exact translations of them. In contrast, using comparable corpora for query translation could have the effect of query expansion, in which not only we could find translations of a query term, but also we could extract related terms to the query terms which are very useful for finding relevant documents to the query. Therefore, we can employ all available resources for finding translation of a query and constructing the query in target language. The main problem of employing several resources for query translation is that each of them may suggest different translations for a query term and determining the correct translation from the candidate translations is difficult. Also, the accuracy of different translation resources for query translation is not the same and using simple methods for fusing translation resources such as linear combination will not lead to a good performance.

Usually in CLIR, queries are considered as bags of words and each query word is considered to be independent of other query words. Assuming query words to be independent is not realistic. In fact, some query words may have different senses in different situations. Using other query words, we can detect the correct sense of ambiguous words in the query. Therefore, the context information should be considered in query translation process to achieve a good performance. However, how to exploit contextual information in query translation process is another challenge.

In this paper, we propose a method for exploiting different translation resources in query translation and considering contextual information in query translation process. This method uses Learning to Rank (LTR) approach for exploiting different translation resources in order to find translations of query terms more accurately. LTR is the task of ranking objects regarding to a query. LTR is originally proposed for information retrieval in which the goal is ranking documents in response to a given query. However, recently, LTR approach has been widely used in other applications [3]. In this paper, we map the problem of finding translations of a query term to a ranking task and use LTR approach for ranking the translations of the query term. We consider each query term as a query and candidate translations of the query term as documents. By this view, the problem of finding accurate translations of a query term is equal to the problem of finding relevant documents to a given query. After mapping the query translation task to a ranking task, we use LTR approach for ranking the candidate translations and finding the correct translations. For constructing a ranking model using LTR, we need training data, containing a set of source language words and their corresponding candidate translations with the ground truths. LTR approach uses different features for learning a ranking model. Therefore, we should define proper features to construct the ranking model. In this paper, we define different features using different translation resources. In fact, the proposed LTR approach uses information extracted from different translation resources for finding accurate translations of a query term. The features used in this research could be categorized into two categories: translation relation based features and context based features. Translation relation based features are based on the translation relations that are extracted from different resources for each pair of source and target language words, such as the translation probability. For extracting context based features, we use the contextual information in translation resources such as the point-wise mutual information of the target language candidate word with the source language query terms.

Concisely, the main contributions of this paper are as follows:

- Using LTR for query translation: In this research, we apply the LTR approach for translating queries in CLIR task. Previous works used translation resources for translating queries and after translating queries they extracted features from query-document pairs and employed LTR methods for constructing a ranking model using constructed training data. Unlike previous works, we use translation resources for extracting features and constructing training data. Using training data, we construct a translation ranking model and employ it for translating queries.

- Exploiting different translation resources for CLIR: Although different studies have been done for exploiting multiple translation resources for CLIR, most of them used a simple approach such as linear combination for exploiting them. This paper proposes an automatic approach for exploiting different translation resources based on LTR approach in query translation process.

- Defining and employing different features for constructing translation ranking model: We define a wide variety of features and use them for constructing the ranking model. Our results in Section V, show the effectiveness of defined features in translating queries accurately.

We do the CLIR task of CLEF 2008 and 2009: "Retrieving Persian documents from queries in English" for evaluating the proposed method. We employ the proposed method for translating the English query words and utilize the translated queries to retrieve Persian documents. We use two parallel corpora: TEP [4] and 20M, UTPECC comparable corpus version 2.0 [5] and a bilingual English-Persian dictionary as the resources of extracting features. Also, we use Wikipedia parallel corpus [6] for constructing the training data and labeling the translation candidates. After constructing the ranking model using LTR approach, we employ the learned model to translate English queries and construct queries in Persian. We use the constructed Persian queries for retrieving documents in Persian. We employ Hamshahri collection [7], which is used in CLEF 2008 and 2009 as test collection. This corpus contains English queries, corresponding Persian queries, and about 166,000 documents in Persian. The results show that using LTR approach for query translation significantly outperforms all single resource based CLIR methods. Also, our results show that the proposed methods outperform the linear combination method which is one of the most used methods for translation resource combination in CLIR. We analyze the impact of different features in constructing the ranking model. Our results show that translation relation based features alongside context based features resolve most of main problems of single resource based CLIR methods.

The remainder of this paper is organized as follows. In Section II, we review the previous work on CLIR and using LTR for CLIR. In Section III, we describe the proposed LTR approach for query translation. Section IV explains the features used in this paper. Section V explains the design of experiments and the results of different CLIR methods. Also, in this section, we discuss impact of different features on constructing the LTR based translation model and the effect of size and quality of corpora on the accuracy of CLIR. Finally, Section VI concludes the paper and describes the future work.

## II. Related work

Different translation resources have been used for query translation in CLIR. In fact, each resource that could provide a translation relation between source and target language words could be employed for query translation. Four translation resources have been widely used in CLIR for query translation [2]: 1) Dictionaries 2) Comparable corpora 3) Parallel corpora and 4) Machine translators. In this research, we did not employ machine translators for query translation because machine translators usually are used for translating a complete sentence and using them for translating queries, which are usually a set of keywords, will not have good performance in CLIR. The mentioned translation resources usually provide term to term matching between source and target language words. However, by using them, the meaning of the query in semantic level will not be considered. Consequently, some problems such as ambiguity remain in translation process, which could have a bad effect on the performance of CLIR system. To overcome this problem, in some researches, approaches such as word sense disambiguation have been utilized while in other researches semantically annotated resources such as ontologies in query translation process have been employed.

Among different translation resources, parallel corpora have been used more than others in CLIR [1, 2]. Using parallel corpora, we can extract translation relations between source and target language words. Different methods have been employed for extracting translations from parallel corpora. Among these methods, IBM model-1[8] is the most used method for translation knowledge extraction. This method provides a probabilistic mapping between source language words to target language words and vice versa. In fact, the output of this method is two probabilistic lexicons, in which for each pair of source language word, $e$, and target language word, $f$, probabilities of translations $p(f|e)$ and $p(e|f)$ are provided. Using these probabilistic lexicons, we can translate each query term to target language and after translating all query terms, we can construct the query in target language. After constructing the query in target language, the traditional IR methods such as Okapi BM25 method could be employed for retrieving documents [9]. Also, the translation probabilities could be employed in language model based IR approaches for calculating the relatedness scores of documents to queries. Berger and Lafferty [10] used the translation probabilities for estimating the language model of query in target language. After estimating the query language model in target language, they used language model based IR methods for retrieving documents in target language. Similar research used this approach for CLIR such as the research done in [11]. Also, in another study, Lavrenko et al. [12] used language modeling approach for CLIR. However, instead of using IBM model-1 for extracting translation knowledge from parallel corpora, they directly estimated the probability of relevance of each target language word with regard to a given source language query and calculated the language model of query in target language.

Comparable corpora are other useful resources for query translation in CLIR. After extracting translation knowledge from a comparable corpus, the methods described for parallel corpora based CLIR could be employed for comparable corpora based CLIR. Different approaches have been proposed for extracting translation from comparable corpora ([13]-[15]). The main intuition behind these methods is that a pair of source and target language words that usually co-occur in aligned documents are more likely to be translations of each other. Using this intuition, Tao and Zhai [13] estimated the associations between source and target language words. They estimated the probability distribution of words in documents of comparable corpora and considered the source and target language words that have similar probability distributions to be translations of each other. Talvensaari et al. [14] used similar approach for extracting translation knowledge from comparable corpora. In addition to co-occurrence information, they used similarity scores of aligned documents for calculating the association score of pairs of source-target language words. Rahimi and Shakery [15] proposed a method based on language modeling framework for extracting translations from comparable corpora. They calculated the contribution of words in constructing the documents of the comparable corpus. Then, they considered two words in two languages that have similar contributions in constructing aligned documents to be translations of each other.

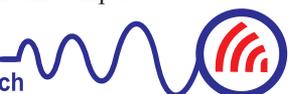

One of the main problems of word to word query translation is existing ambiguity in query words which word to word translation could not resolve it. In fact, a query word could have different meanings (senses) in target language and finding the correct translation of the query word is very challenging in CLIR. To solve this problem, different methods([16]-[19])have been proposed. The main idea behind these methods is using the co-occurrence information of translated query words for finding the correct translations. In fact, these methods make the assumption that the correct translation of a query word should have high co-occurrence with translations of other query words. In this research, we used this idea for defining different features and finding the correct translations of query words.

Some CLIR methods utilize semantic and lexical relations between words and concepts to increase the accuracy of CLIR. The main intuition behind these methods is bridging the gap between words and their meanings. These methods add context to query and expand the query. Adding context to query could resolve the problem of ambiguity. Ontologies and thesauri are potential resources for extracting semantic level information. These resources have been widely exploited in CLIR. EuroWordNet[20] is a semantic network of words of seven European languages in which the words of different languages are connected with inter-lingual links. In many researches([21]-[23]), this ontology has been used for matching queries and documents. For example, instead of indexing words, Gonzalo et al. [21] used EuroWordNet's Inter-Lingual-Indexes as index units. Using this approach, the process of CLIR becomes language independent. In fact, each word (sense of word) is indexed with its corresponding word (sense of word) in other languages. Therefore, query in any language could be matched with documents in any language.

In some researches ([24]-[26]), ontologies are used beside other resources such as dictionary for adding semantic level information to the translations and finding the correct meanings of words. For example, Pourmahmoud and Shamsfard[24] employed a dictionary for query translation and an ontology for expanding query with related words (phrases). The results of this research show that using ontologies and semantic level resources for query translation could improve the accuracy of CLIR.

Other researches ([27]-[30]) also tried to consider the meanings of words in semantic level in query translation. For example, some researches ([29], [30]) used Latent Semantic Indexing (LSI) for query translation and mapped the queries and documents into a multilingual space. In general, using semantic level resources alongside other resources could help to achieve good accuracy in CLIR. However, using them together and combining them is very difficult because originally the translation units proposed by each resource are different. In this paper, we use information extracted from parallel corpora, comparable corpora and dictionary for translation. Some of the information is based on the co-occurrence of words which could play the role of considering semantic information in query translation process.

In order to increase the accuracy of CLIR, multiple resources could be exploited simultaneously. Different methods are proposed for using multiple translation resources ([31]-[36]). The main intuition behind these methods is that each resource has some advantages for query translation and by using all of them, we can exploit the advantages of all resources and translate queries more accurately. Among these methods, the method of Kadri and Nie [34] is similar to our method. They extracted different features using different resources. Using the extracted features they employed a neural network for estimating the weights of different features and different translation resources in order to combine them for extracting the translations of query words. We used a similar approach, but instead of using neural networks for estimating the weights of different resources, we used LTR approach. Also, in addition to some of the features used in [34], we defined new features and employed them to construct a translation ranking model.

Recently, LTR approach has been widely used in different applications of IR and natural language processing [3].Multi Lingual IR (MLIR) and CLIR are applications of IR that LTR approach have been employed in them. MLIR is applying CLIR on more than two languages and so CLIR is a special case of MLIR. Different studies have been done on using LTR in MLIR such as using LTR for merging the lists in MLIR [37] and learning a scoring function from multilingual parallel corpus for ranking documents [38]. LTR approach also is employed in CLIR. Azarbonyad et al. [39] used LTR approach for constructing a ranking model in parallel corpora based CLIR. They utilized parallel corpora for mapping monolingual features to cross lingual ones and used the mapped features for constructing a ranking model. In fact, they used a parallel corpus for estimating cross lingual features and learning a ranking model. After constructing the ranking model, Azarbonyad et al. [39] directly employed the model for ranking documents. Unlike the method proposed in [39], which employed parallel corpora for mapping monolingual features to cross lingual equivalents without query translation and constructed an LTR based learning model, we used translation resources for extracting different features and employed LTR approach for translating queries.

III. LEARNING TO RANK FOR QUERY TRANSLATION

Recently, LTR methods have been widely used in different applications of information retrieval. The main focus of LTR methods is using several features for ranking documents in response to a query in information retrieval task. Although LTR methods are originally proposed for document retrieval task, we could use these methods in other applications such as query translation. In this paper, we apply LTR approach for query translation in CLIR task. The main step of applying LTR methods in other applications is mapping the application to a ranking problem. In query translation task, we map the problem of finding translations of a query term to the ranking problem. We consider each query term as a query and the translation candidates of the query term as documents. In this way, our problem could be viewed as a ranking problem.

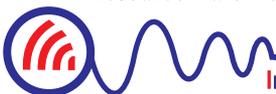

LTR methods need training data for constructing the ranking model. In IR task, the data is a number of queries, a list of judged documents for each query, and a relevancy label for each pair of query-document which determines whether the document is relevant to query or not. We need similar training data for constructing the translation ranking model. We consider each query word as a query and its translation candidates as documents. Therefore, in training data, we should have a number of source language words, their translation candidates, and the labels of them. The label for each candidate translation of a source language word determines whether the candidate translation is exact translation of the source language word or not. Finding accurate label of pairs of words is very difficult and needs human judgment. We use several translation resources for finding translation relations of source and target language words.

If all translation resources agree that two given words are translations of each other, it is likely that such words be exact translations of each other. However, using only translation relations without any attention to the context around the words is not enough for estimating the translation labels of words. For example, the translation of "cup" could be either "جام/cup/jam" (in x/y/z notation, x is the word in Persian, y is translation of x in English, and z is the pronunciation of x in English) or "فنجان/cup/fenjan" in Persian. If the context around the "cup" is about sport, the first translation will be true. For example, if the sentence is "Brazil hosts football world cup", the translation of "cup" will be "جام/cup/jam". In other situations, the translation of "cup" can be "فنجان/cup/fenjan". Therefore, determining the exact translation of a word regarding to a context is beyond using only dictionaries or lexicons. Due to this issue, we use another resource for determining the contexts of words and translating them more accurately. To do so, we use a parallel corpus for labeling translation candidates. The process of acquiring the labels of candidate translation regarding to a source language word is shown by as example in Fig. 1.At first, we use IBM model-1 for finding the alignment of words in sentences. After using IBM model-1, we have a word aligned parallel corpus, in which each word of a sentence in source language is aligned to its corresponding word in corresponding aligned sentence. For example, in the Fig. 1, the word "cup" is appeared in sentence "Brazil hosts football world cup" and the corresponding aligned sentence in Persian is "برزیل میزبان جام جهانی فوتبال است", so it is more likely that "cup" is aligned to "جام/cup/jam". After constructing word aligned parallel corpus, the candidate translations are extracted. For example, in Fig. 1, we can see that from one sentence, five translation pairs are extracted. After extracting translation candidates, we use lexicons extracted from translation resources for validating translation pairs and labeling them. Using the lexicons of translation resources and the word aligned parallel corpus, we can estimate the labels of translations accurately. We use this fact that if two words in target and source languages are accurately aligned in a sentence of parallel corpus, then in lexicons that are extracted from different resources these words should have a translation relation. In fact, if a source language word is aligned to a target language word in a sentence, the target language word should be in translations of the source word in other lexicons. For example, in Fig. 1, we can see that for the word "world", translation extracted from word aligned corpus is "جهانی/world/jahani". Also, we can see that in lexicons that are extracted from translation resources, the words "world" and "جهانی/world/jahani" have translation relation. Therefore, we label the word "جهانی/world/jahani" as exact translation of the word "world". In fact, the translation of the word "world" in the context of "Brazil hosts football world cup" in Persian is "جهانی/world/jahani". We label other translation candidates that are extracted from lexicons of different resources zero, which means that they are not exact translation of "world" in the context. We use two parallel corpora, a comparable corpus, and a bilingual dictionary for query translation. Also, we use another parallel corpus for finding the alignment of words in sentences. After finding alignments of words in parallel corpus, we use other translation resources for validating the alignments and labeling the translations. For this purpose, we label each candidate as true when source and target language words have a translation relation in lexicons extracted from other parallel corpora and comparable corpus.

The general process of LTR for translation could be summarized as follows: Suppose that we have two different sets. We name these sets as source language words $SW = \{sw_1, sw_2, …, sw_m\}$ and target language words $TW = \{tw_1, tw_2, …, tw_n\}$. Given a member $sw_i$ of $SW$ and a subset $tw$ of $TW$ ($tw$ is the set of candidate translations of $sw_i$), we rank the members of $tw$ based on the information we get from $sw_i$ and $tw$. We perform the process of ranking using a ranking (scoring) function:

$$F(sw_i, tw) : SW * TW^k \to R^k, \qquad (1)$$

where $k$ is the size of $tw$ and $R$ is the set of real words. In fact, the scoring function gives scores to the members of tw: $S_{tw} = F(sw_i, tw)$. Then, the translation candidates are ranked according to these scores: $R = sort_{S_{tw}}(tw)$. Instead of calculating scores for all members of $tw$ and ranking them, we calculate the score for each member of $tw$. In fact, in this paper instead of using $F(sw_i, tw)$, we use a function $f(sw_i, tw_j)$ which calculates the score of a member of $tw$ regarding to $sw_i$. Therefore, the ranking actually will be performed using function $f$:

$$f(sw_i, tw_j) : SW * TW \to R \quad (2)$$

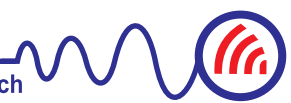

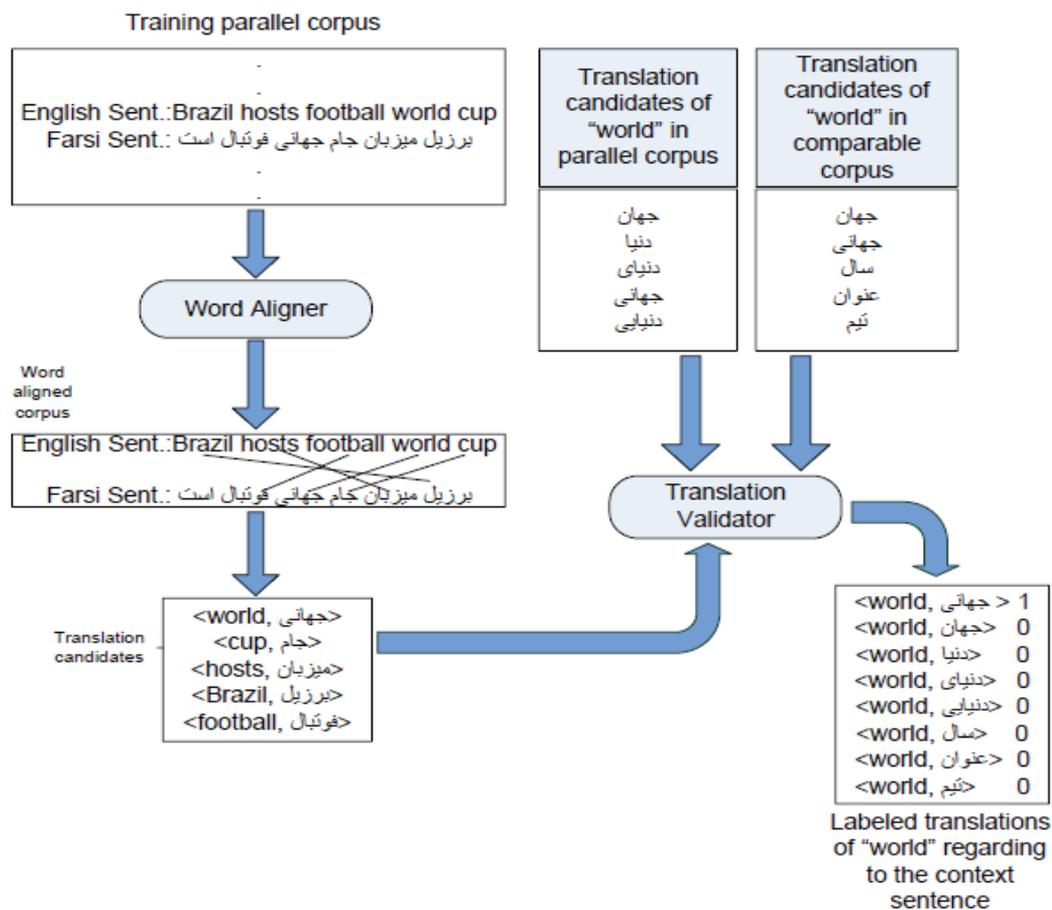

Figure 1. The steps of labeling translation candidates

Thus, the scoring and ranking functions will be as follows: $s_{tw_j} = f(sw_i, tw_j)$ and $Ranking = \forall s_{tw_j} \in tw \; sort_{s_{tw_j}}(tw)$.

The proposed method has three main steps: constructing the training data, learning a ranking model, and using the ranking model for ranking translations. The first step is described above. In second step, we use the training data for learning a ranking model. The training data consists of source language words and their associated translation candidates with their labels. Using these data, the learning machine constructs a ranking model. In fact, the goal of this step is constructing the function $f(sw_i, tw_j)$. The scoring function is actually a combination of different features extracted in the first step.

In the third step, we use the ranking model for scoring and ranking the translation candidates of a new word. In fact, given a new source language word $sw_{m+1}$, we rank the translations candidates of $sw_{m+1}$ and use the top ranked candidates as translations of $sw_{m+1}$.

Also, the main steps of our method for query translation are shown in Fig. 2. First we construct the training data, which contains the source language words and their translation candidates. The process of constructing the training data is shown in Fig. 1. In Fig. 2, $<T_i, S_j>$ represents a source language word $T_i$ in a source language sentence $S_j$. For each $<T_i, S_j>$ we have its candidate translation and labels of them. $TRS_{ik}$ corresponds to the $k^{th}$ translation candidate of the word $T_i$ and $L_{ik}$ shows the label of the candidate translation regarding to the source language word. $L_{ik}$ will be one if the candidate translation $TRS_{ik}$ is exact translation of source language word $T_i$, zero otherwise.

After labeling pairs of source and target language words, we extract different features for these pairs using different translation resources. The features are described in Section IV. After extracting features from pairs of source and target language words we use the features for learning a ranking model. Training data is the input of learning system. Using this data, the learning system constructs a ranking model. The ranking model (translation model) could be used for scoring the candidate translations. Using features extracted from query words and their candidate translations and by employing the ranking model, we can score the candidate translations of query words and re-rank them for finding the exact translations of query words. The output of our system will be re-ranked translation candidates of query words. We can employ the re-ranked translations for constructing the query in target language and retrieving documents.

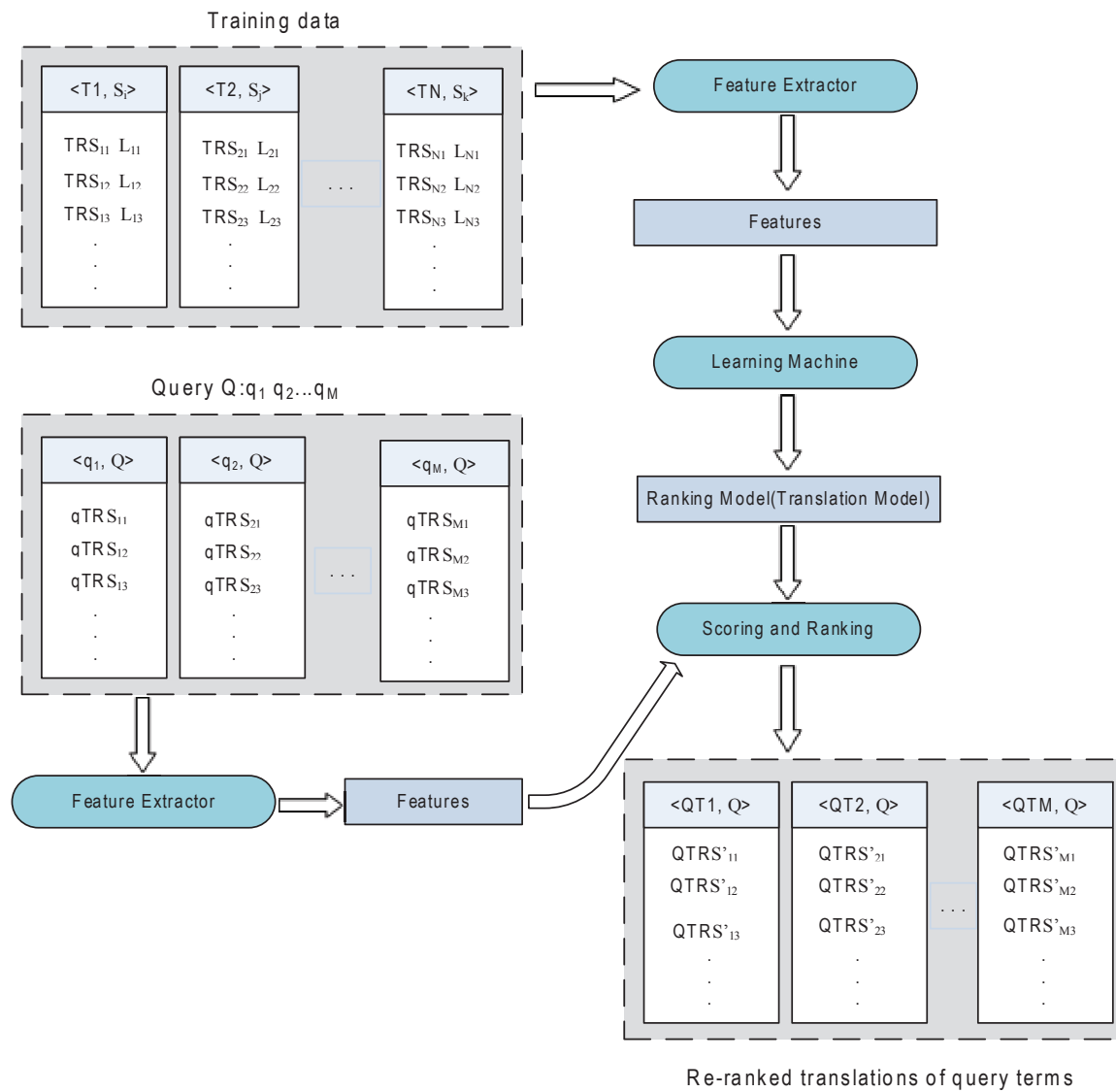

Figure 1.  The steps of applying LTR for query translation in CLIR

## IV. TRANSLATION FEATURES

In this section, we describe the features used for constructing the translation ranking model. We extract a wide variety of features using different translation resources for each pair of source and target language words. The features used in this research could be categorized to two categories: translation relation based features and contextual information based features.

### A. Translation relation based features

The features of this category are estimated using translation relations extracted from different translation resources. In this section, we describe the features of this category.

**Translation probability:** This feature is the probability of translating the source language word to the candidate target language word. For extracting this feature from parallel corpora, we use IBM model-1. In order to extract translation probabilities from comparable corpora, we use Rahimi and Shakery [15] method, which is one of the best performing methods for extracting translation relations from comparable corpora. Unlike most of other methods of extracting translation relations from comparable corpora which use only alignments between source and target language documents, Rahimi and Shakery [15] method uses alignment scores of documents for extracting translation relations. Since dictionary does not contain translation probabilities, we simply consider translation probabilities to be uniform, i.e. if an English word has $N$ translation candidates, we will consider the translation probability of each candidate to be equal to $1/N$.

**Reverse translation probability:** Another feature we employ is the reverse translation probability. We use this fact that if two words are translations of each other, both the probability of translating target language word to source language word and probability of translating source language word to target language word should be high. Again, we use IBM model-1 for extracting reverse probabilistic dictionary from parallel corpora. Also, we use Rahimi and Shakery [15] method for extracting reverse probabilistic dictionary from comparable corpora.

**Translation ranking:** Another feature we employed in this research, is the ranking of the translation candidates of the source word in the candidate translation list of the source word. To extract this feature, we sort the candidate translations of source words by their translation probabilities and consider the rank of the candidate target word as the feature.

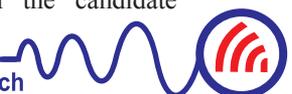

**Translation probability difference:** The difference between the translation probability of a candidate translation and the highest translation probability in translation candidates of the source word is another feature that we use in this paper.

In the proposed method, we also use some features that are independently extracted for source and target words. Term frequency and inverse document frequency of candidate translations in different translation resources are the features of this category.

**Term frequency:** We use the frequency of source word in each resource as a feature. If frequency of a word in a resource is low, we have less confidence in the translations that the resource provides for the word. Therefore, the frequency of word in the resources could be a good feature for selecting the translation resource. Similarly, frequency of target language words could be a good feature for determining the correct translations.

**Inverse document frequency:** Although the frequency of a word in different resources could be a good feature for selecting the correct resource, if a word excessively is repeated in are source, it could be a general word. Usually, the general words, which are frequent in a resource, have incorrect translation relations with many words. We should prevent these words in translation process. For example our observations show that the Persian word "سال/year/sal" has translation relation with most of English words in the lexicon extracted from comparable corpus. To solve this problem, we use Inverse Document Frequency (IDF) of words as another feature. Usually, general words have low IDF's, so using this feature we can prevent facing the general words problem.

**Entropy of words:** Another feature that we use to solve the problem of general words is the entropy of target words in the lexicons extracted from different resources. The entropy for a given target language word is defined as:

$$E(w_t) = -\sum_s p(w_t|w_s) * \log(p(w_t|w_s)), \quad (3)$$

where $w_t$ is a target language word, $w_s$ is a source language word, and $p(w_t|w_s)$ is the translation probability provided by a resource to the given source and target language words. The sum is taken over the source language words that have a translation relation with the target language word. If the entropy of a given word is high, it will be more likely that the word is a general word. Similarly, we use reverse translation probabilities for estimating the entropy of source words. Using each of resources, we extract these features and we have two entropy based feature regarding to each resource.

*B. Context based features*

To consider contextual information in translation process, we define different features. In this section, we describe the features of this category.

**Point-wise mutual information:** We use Point-wise Mutual Information (PMI) for considering the correlation of words in translation process. PMI is one of the common measures of estimating the coherence of two words in a corpus. PMI is defined as:

$$PMI(w_1, w_2) = \log\left(\frac{p_{12}}{p_1 * p_2}\right), \quad (4)$$

where $p_{12}$ is the probability of occurrence of words $w_1$ and $w_2$ simultaneously in a corpus and defined as:

$$p_{12} = \frac{MutualCount(w_1, w_2)}{N}, \quad (5)$$

Where $MutualCount(w_1, w_2)$ is the number of documents that contain both $w_1$ and $w_2$ and $N$ is the total number of documents in the corpus. $p_1$ and $p_2$ are the probability of occurrence of $w_1$ and $w_2$ respectively. $p_i$ is defined as the number of documents containing $w_i$ divided by $N$.

For each word $w$, we calculate the correlation of that word with other words of the sentence that contains $w$ as:

$$C(w, S) = \sum_{w_i \in S, w_i \neq w} PMI(w, w_i), (6)$$

Where $S$ is the sentence containing $w$. $S$ is the sentence of training parallel corpus that we extracted the pair of source-target language words from it. We calculate these features for each pair of source and target language words. In fact, in training data, we have two features from this class: one calculated using the source word and considering $S$ as the source language sentence, and another one calculated using the candidate target language word and considering $S$ as the aligned target language sentence. For calculating this feature, we use comparable and parallel corpora. Using each resource, we extract these features and employ them for constructing the ranking model. Unlike comparable corpora, parallel corpora are aligned in sentence level. Therefore, we consider each sentence of parallel corpus as a document and calculate the counts.

**Cross lingual point-wise mutual information:** Another context based feature is the Cross Lingual PMI (CLPMI) which is the PMI of pair of source and target language words. For extracting CLPMI we use alignments in different resources. Since source and target language words are in different languages, we cannot calculate the *MutualCount* like the *MutualCount* of PMI. We calculate the CLPMI's *MutualCount* as follow:

$$MutualCount(w_s, w_t) = \sum_{\substack{A_{ij} \in A \\ w_s \in d_i \\ w_t \in d_j}} 1, \quad (7)$$

Where $A$ is all alignments, $A_{ij}$ is an alignment between document $d_i$ in source language and document $d_j$ in target language.

Using CLPMI, we calculate total PMI of each source language word with the sentence, $S_t$, in target language as follow:

$$C(w_s, S_t) = \sum_{w_i \in S_t} CLPMI(w_s, w_i) \quad (8)$$

Similarly, we calculate the total CLPMI of each target language word with the source language sentence and use it as a feature in constructing the ranking model.

**Number of relevant words:** Another context based feature is the number of source language words in source sentence that have a translation relation with the candidate target language word. As the value of this feature for a target language word increases, the probability of translation of the target word to the source word should be increased.

## V. EXPERIMEN RESULTS

In this section, we evaluate our methods to show their effectiveness. In order to evaluate different methods, we use Hamshahri collection and do the CLEF 2008 and 2009 tasks: retrieving Persian documents from queries in English. At first, we describe the datasets used in this research for evaluating the proposed method and then we present the results of different methods.

### A. Datastes

For evaluating different CLIR methods, we use CLEF Hamshahri collection. This collection contains 166,744 documents in Persian and 100 queries in Persian and English languages. We use the title of English queries for retrieving Persian documents.

We employ three different parallel corpora for implementing the proposed method. Wikipedia parallel corpus is used for extracting translation candidates and their labels. This corpus consists of 282,853 aligned sentences in Persian and English. The size of this corpus is 2,299,025 words in Persian and 2,288,807 words in English. We use TEP and 20M parallel corpora as translation resources for extracting features. TEP is constructed from subtitles of about 1,600 movies. This corpus contains 612,086 aligned sentences which are about 3,783,720 words in Persian and 3,893,249 words in English. The 20M corpus is a combination of four different parallel corpora: Roman parallel corpus [40], ELRA Persian-English parallel corpus [41], a part of Mizan parallel corpus [42], and ITRC parallel corpus [43]. The size of this corpus is about 1,109,584 sentences. This corpus contains 19,779,899 Persian words and 19,848,527 English words.

We use UTPECC comparable corpus version 2.0 for extracting translations of words. This corpus is constructed from aligned news articles in English and Persian. About 10,724BBC news articles are used for constructing the English side of this corpus and about 5,544Hamshahri news is used for constructing the Persian side and the total number of alignments between English and Persian documents is 14,979.

### B. Experimental results

We use Wikipedia parallel corpus for extracting translation candidates. After extracting translation candidates, we employ different resources for calculating features and constructing the training data. We use LTR methods for constructing a ranking model by employing the train data. In calculating some of the features, we assume that the candidate translations are in a sentence and we use the sentence for estimating some of features. In query time, we do not have any sentence in target language, but we can consider the query in source language as a sentence for estimating source language related context based features. To solve the problem of lack of sentence in target language, we use the top five translations of each source language query term and construct the sentence in target language. In fact, in target language sentence, for each source language term, we have five translation candidates. The top five translations are extracted using 20M corpus because this corpus has the best performance in CLIR compared to other resources.

We do English-Persian CLIR task using different translation resources. We used different LTR methods for constructing the translation model (AdaRank [44], IR SVM [45], and Coordinate Ascent [46]). Among them, IR SVM had the best performance. Therefore, we only report the results of this method.

In Table I, the results of different CLIR methods and the results of monolingual IR are shown. For monolingual IR, we used Persian queries for retrieving Persian documents. We employ Okapi BM25 retrieval method for scoring and retrieving documents.

In dictionary based CLIR, we employed top $N$ translations of each query word for constructing the query in target language. After constructing the query in target language, we use Okapi BM25 method for retrieving documents in target language. We achieved the best performance where $N=6$. Therefore, we only reported the results of dictionary based CLIR with $N=6$. The best performance of dictionary based CLIR in terms of MAP is 0.1121, which is 0.25 of MAP of monolingual method. As the results show, dictionary is not a good resource for query translation. In fact, dictionary has many shortcomings for query translation such as limited coverage and lack of named entities. Our results show that using dictionary, we have 64 Out-Of-Vocabulary (OOV) words. The OOV words usually are named entities, which are very important in information retrieval.

Like dictionary based CLIR, in comparable corpus based CLIR, we employ top $N$ translations of each query word for constructing query in target language. We achieved the best performance when $N=3$. Therefore, we report the results of comparable corpus based CLIR for $N=3$. As the results show, the performance of comparable corpus based method is better than dictionary based method. The performance of comparable corpus based CLIR in terms of MAP is 0.1545, which is 0.37 of MAP of monolingual IR. In comparable corpus based CLIR we have 24 OOV words, which is lower than dictionary based CLIR. In fact, we expected the coverage of comparable corpus to be better than dictionary.

The results of parallel corpora based method are also shown in Table I. As the results show, parallel corpora based CLIR method has the best performance comparing to other methods. We used top $N$ translations of each query word for constructing query in target language. The best results of parallel corpora based method are achieved when $N=5$ so, we only report the results of parallel corpus based CLIR for

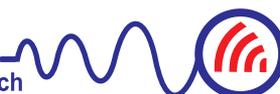

$N=5$. Using parallel corpora as translation resource, we employed IBM model-1 and extracted probabilistic lexicon. The performance of parallel corpus based method, when we use TEP corpus, in terms of MAP is 0.2652, which is 0.59 of MAP of monolingual method. Using this resource we have 13 OOV words, which is lower than comparable corpus based method and shows its effect on the results. The 20M corpus is even better than TEP corpus. Using this resource the performance of CLIR in terms of MAP is 0.3064, which is 0.74 of MAP of monolingual method. Using this resource we only have 4 OOV words. This shows that 20M corpus has good coverage.

In this paper, we implement the linear combination method for combining different translation resources. This method assigns weights for translation resources based on the accuracy of them and uses the weighted sum of translation probabilities extracted from different resources. This method estimates the translation probability of a target language word $f$ to a source language word $e$ as follows:

$$P(f|e) = \lambda_i P_{R_i}(f|e), \qquad (9)$$

where $P_{R_i}(f|e)$ is the translation probability extracted using resource $R_i$ and $\lambda_i$ is the weight of $R_i$. After estimating translation probabilities using Equation 9, we use top $N$ translations that have highest translation probabilities for each English query word. In this paper, we set $N=5$. We examined different values for $\lambda$'s. The best performance in terms of MAP is achieved when we set the $\lambda = 0.7$ for 20M parallel corpus, $\lambda = 0.2$ for TEP parallel corpus, $\lambda = 0.1$ for UTPECC comparable corpus, and $\lambda = 0$ for dictionary. In fact, our experiments show that using linear combination, dictionary does not have any contribution in improving the accuracy of CLIR. The best result of linear combination in terms of MAP is 0.3106. This result shows that linear combination improves the CLIR accuracy by 1%.

The results of using LTR for query translation are shown in the last row of Table I. In this method, we first used LTR method and learned a translation ranking model. Then, we employed top $N$ translations of each query word after re-ranking for constructing query in target language. We achieved the best performance when $N=5$ so, we only report the results for $N=5$. Finally, using the constructed query, we implemented Okapi BM25 retrieval method for scoring and retrieving documents. The results show that using LTR for query translation improves the accuracy of CLIR. The best performance of LTR based CLIR in terms of MAP is 0.3217 which is 0.78 of MAP of monolingual method. This result is also 0.05 better than the result of 20M corpus based CLIR, which is the best result of single resource based CLIR. Also, the results show that LTR based translation combination method outperforms the linear combination method and the performance of this method is 0.04 better than the performance of linear combination method in terms of MAP. We conducted statistical significant test (t-test) on the improvements of LTR based CLIR over single resource based CLIR methods and the linear combination method. Our results show that all improvements in terms of MAP are statistically significant (p-value < 0.005).

In Table II, top five translations of each query term of query: "2002 world cup" obtained from different translation resources and LTR method are shown. This query contains a phrase "world cup" and the words "world" and "cup" have different meanings in Persian. However, as a phrase, the right translation of the English phrase in Persian is "جام جهانى". In fact, without paying attention to the other words of the query, we cannot translate each of words. As can be seen from Table II, using each of single resources, we cannot translate this query accurately. In fact, each resource suggests the most frequent translations of the source word as translation of the word. The most frequent translation of word "world" in Persian is "جهان/world/jahan" and the most frequent translation of word "cup" is "فنجان/cup/fenjan". Thus, using only the translation probabilities, at least we cannot translate queries that contain phrases accurately. As can be seen from Table II, LTR based translation method overcomes this problem. In LTR based method, some candidates get a negative score. We ignore the candidates that have negative score and normalize the scores of other translation candidates to be in [0,1] interval. The scores of candidates after normalization also are shown in Table II. As can be seen, LTR based translation method translates the words "world" and "cup" accurately. This is the effect of context based features which we defined in Section IV. Also, LTR method uses different features extracted from different translation resources. As can be seen from Table II, comparable corpus translates the words "world" and "cup" accurately, but it does not translate the world "2002" accurately. In contrast, other translation resources translate the word "2002" accurately. LTR method exploits all translation resources and translates the words of query more accurately.

*C. Discussions*

In this section, we discuss two important issues that could have high impact on the performance of LTR based translation method. One of these issues is the effect of size and quality of translation resources in CLIR and another issue is the impact of different features in constructing the translation candidate ranking model.

*1) The effect of size and quality of translation resources in CLIR*

One of the important factors that affects the performance of CLIR is the quality of translation resource. In this section, we study the effect of size and quality of different corpora on the accuracy of CLIR. In Table III, statistics of different parallel corpora are shown. Also, in Table IV, the results of using different parallel corpora for CLIR are shown. 20M corpus has the biggest size among parallel corpora. Also, this corpus has the best performance in CLIR. This corpus has more unique words compared to other corpora. Thus, this corpus has good coverage and could translate most of words. Among other corpora, Roman corpus has the biggest size. However, this corpus does not have good performance in CLIR. Roman corpus is constructed from English-Persian books. Usually, books cover a special range and they have limited coverage. From Table III we can see that the number of unique words of Roman corpus is lower than other corpora. Therefore, using this corpus, we cannot translate queries accurately. Although the size of TEP corpus is smaller than Roman corpus, it has better performance in CLIR. TEP corpus is constructed from about 1,600 movie subtitles. Since the number of movies used for constructing this corpus is high, this corpus has good coverage. Also, the sentences of this corpus are shorter than sentences of other corpora. The average sentence length of this corpus is about 6.2 words, while the average sentence length for Roman corpus is about 15.3 words. Usually, the probabilistic dictionary extracted from a corpus with short sentences has better accuracy from the one extracted from a corpus with big sentences. The main problem of TEP corpus is that it is very informal because the conversations in movies are informal. The translation extracted from this corpus for queries are also informal, while the documents of Hamshahri corpuses are not informal. Therefore, one of the reasons that this corpus has lower performance compared to 20M and Wikipedia corpora is its informality. The Wikipedia corpus is smaller than other corpora. However, it has good performance in CLIR. This corpus is constructed from Wikipedia articles, which are in different domains. Therefore, this corpus has good coverage. Additionally, the sentences of this corpus are short. The average sentence length of this corpus is about 8.2.

TABLE I.    RESULTS OF DIFFERENT CLIR METHODS

| Method | MAP | %Mono | P@5 | %Mono | P@10 | %Mono |
|---|---|---|---|---|---|---|
| Monolingual | 0.4126 | - | 0.702 | - | 0.643 | - |
| Dictionary based method | 0.1121 | 25 | 0.198 | 25 | 0.183 | 27 |
| Comparable corpus based method | 0.1545 | 37 | 0.306 | 44 | 0.225 | 35 |
| Parallel corpus (TEP) based method | 0.2652 | 59 | 0.44 | 55 | 0.41 | 60 |
| Parallel corpus (20M) based method | 0.3064 | 74 | 0.524 | 75 | 0.499 | 78 |
| Linear combination method | 0.3106 | 75 | 0.538 | 77 | 0.508 | 79 |
| LTR based combinational method | 0.3217 | 78 | 0.562 | 80 | 0.518 | 81 |

TABLE II.    TRANSLATIONS EXTRACTED FROM DIFFERENT RESOURCES FOR QUERY: "2002 WORLD CUP"

| Translation resource | Top five translations of query terms |
|---|---|
| Dictionary | 2002: OOV<br>world: جهان (world), دنیا (world), گیتی (world), عالم (world), روزگار (period)<br>Cup: فنجان (cup), پیاله (cup), جام (cup), ساغر (cup) |
| UTPECC Comparable corpus | 2002: سال (year) 0.11, درصد (percent) 0.06, 2002 (2002) 0.05, اقتصادی (economic) 0.05, مخل (intruder) 0.04<br>World: جهان (world) 0.11, جهانی (world) 0.11, سال (year) 0.04, عنوان (title) 0.03, تیم (team) 0.02<br>Cup: جام (cup) 0.22, فوتبال (football) 0.08, تیم (team) 0.01, بازیکن (player) 0.009, بازیکنان (players) 0.001 |
| TEP parallel corpus | 2002: 2002 (2002) 0.5, سال (year) 0.5<br>World: دنیا (world) 0.58, دنیای (world) 0.19, جهان (world) 0.16, دنیایی (world) 0.03, جهانی (world) 0.03<br>Cup: فنجون (cup) 0.35, فنجان (cup) 0.29, قهوه (coffee) 0.13, استنلی (Stanley) 0.13, لیوان (glass) 0.1 |
| 20M corpus | 2002: 2002 (2002) 0.67, سال (year) 0.30, سالهای (years) 0.01, الاوسط (awsat) 0.01<br>World: جهان (world) 0.47, دنیا (world) 0.26, دنیای (world) 0.13, جهانی (world) 0.12, دنیایی (world) 0.01<br>Cup: فنجان (cup) 0.46, جام (cup) 0.4, فنجانی (cupping) 0.07, لیوان (glass) 0.03, پیمانه (measure) 0.03 |
| Linear combination method | 2002: 2002 (2002) 0.57, سال (year) 0.32, سالهای (years) 0.007, الاوسط (awsat) 0.007<br>World: جهان (world) 0.37, دنیا (world) 0.29, دنیای (world) 0.12, جهانی (world) 0.1, دنیایی (world) 0.01<br>Cup: فنجان (cup) 0.38, جام (cup) 0.3, فنجون (cup) 0.07, فنجانی (cupping) 0.04, لیوان (glass) 0.04 |
| LTR based combinational method | 2002: 2002 (2002) 1<br>World: جهانی (world) 0.65, جهان (world) 0.35<br>Cup: جام (cup) 1 |

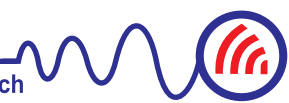

TABLE III. STATISTICS OF DIFFERENT PARALLEL CORPORA

| Corpus | #Bilingual Sentences | #Persian words | #English words | #Unique Persian words | #Unique English words |
|---|---|---|---|---|---|
| TEP | 612,086 | 3,783,720 | 3,893,249 | 114,275 | 73,002 |
| Roman | 399,000 | 6,074,550 | 6,528,241 | 101,114 | 65,123 |
| 20M | 1,109,586 | 19,779,899 | 19,848,527 | 256,549 | 211,544 |
| Wikipedia | 282,853 | 2,299,025 | 2,288,807 | 115,471 | 135,028 |

TABLE IV. RESULTS OF EMPLOYING DIFFERENT PARALLEL CORPORA IN CLIR

| Corpus | MAP | %Mono | P@5 | %Mono | P@10 | %Mono |
|---|---|---|---|---|---|---|
| TEP | 0.2652 | 59 | 0.44 | 55 | 0.41 | 60 |
| Roman | 0.1618 | 39 | 0.324 | 46 | 0.305 | 47 |
| 20M | 0.3064 | 74 | 0.524 | 75 | 0.499 | 78 |
| Wikipedia | 0.2975 | 72 | 0.508 | 72 | 0.472 | 73 |

From the results we can conclude that two aspects of parallel corpora are important in CLIR: 1) coverage of parallel corpora and 2) length of sentences of parallel corpora. Coverage of parallel corpora has a great impact on the performance of CLIR. Since the queries are in different domains, the parallel corpus should have good coverage of different domains. For example, although Roman corpus has greater size than TEP and Wikipedia corpora, its performance are lower than TEP and Wikipedia in CLIR, because it has limited coverage. Length of sentences of a corpus also has a great impact on the quality of probabilistic lexicon extracted from parallel corpus and so on the performance of CLIR. In fact, one of the quality factors of a parallel corpus is the shortness of its sentences. As the length of sentences increase, the accuracy of IBM model-1 in extracting word to word alignments decreases.

*2) Impact of features on constructing the translation ranking model*

In this research, we defined and employed different features using different translation resources. We used these features for constructing a translation ranking model using LTR approach. For analyzing the impact of different features in constructing the ranking model, we use Forward Selection method. We begin with the best performing feature and one by one add other best performing features.

The results show that different features have different contributions in constructing the ranking model. Generally, the translation probabilities extracted from different resources are the most important features in constructing the translation ranking model. The translation probability extracted from 20M corpus using IBM model-1 has greatest impact in constructing the ranking model. Using only this feature, the performance of CLIR in terms of MAP is 0.3064. We name this feature $prob_{20M}$. By adding other features to this feature, we achieve 0.05 improvements in terms of MAP. Among other features, the translation probability extracted from TEP corpus using IBM model-1 has the highest impact on the improvements. We name this feature $prob_{TEP}$. By adding $prob_{TEP}$ to the $prob_{20M}$ the performance of CLIR in terms of MAP increases to 0.3113. This shows that the improvement achieved by adding $prob_{TEP}$ is about 0.02. The cross language PMI of target language word with source sentence, where the PMI calculated using UTPECC corpus is the next best feature. We name this feature $CLPMI_{UTPECC}$. The MAP achieved by adding $CLPMI_{UTPECC}$ to $prob_{20M}$ and $prob_{TEP}$ is 0.3152. Among other features reverse translation probability extracted from 20M corpus using IBM model-1, $CLPMI_{20M}$, $prob_{UTPECC}$, the PMI of target word with target language sentence's words where the PMI is calculated using the 20M corpus are next best features, respectively. Other features also have a few contributions on constructing the translation ranking model.

The performance of CLIR system in terms of MAP when we use only translation relation based features is 0.3144. This shows that adding other translation relation based features to the $prob_{20M}$ feature increases MAP of CLIR by 3%. The MAP achieved by adding context based features to translation relation based features is 0.3217, which shows that context based features further improve MAP by 2%.

The results show that the translation relation based features are more important than context based features. However, context based features also are very useful in constructing the ranking model. The $prob_{20M}$ feature has the best performance in CLIR compared to other features. Also, among other features $prob_{TEP}$ has the best performance. These two features are extracted from two parallel corpora using IBM model-1. In parallel corpus based CLIR, most of researches used IBM model-1 for translating queries. Our results also show that IBM model-1 is the best method for extracting translations from parallel corpus and translating queries. However, IBM model-1 has some shortcomings in order to be employed in CLIR. The main problem of IBM model-1 in translating queries is that it does not consider the context in extracting translations and the translations extracted using this method could have ambiguity. We can solve this problem by adding context based features to the translation based features. The defined context based features in Section IV, solves a big part of shortcomings of translation based features.

## VI. Conclusion and future work

In this paper, we exploited different translation resources for CLIR. To do so, we mapped the problem of finding translations of query words to the ranking problem. After mapping the query translation problem to the ranking problem, we employed Learning to Rank (LTR) approach for constructing a translation ranking model. We used the constructed model for scoring the translation candidates of query words and re-ranked the translation candidates using the scores the model gave to each candidate. We defined different features using different translation resources. In addition to translation relation based features, which are based on the translation information extracted from different resources, we defined different features which use the contextual information contained in different translation resources. Our results show that using LTR for query translation significantly outperforms other CLIR methods. The performance of LTR based CLIR method in terms of MAP is 0.3217, which is 0.05 better than the best performing single resource based CLIR method. This result shows that LTR method exploits different translation resources for CLIR very well. Our results also show that although translation relation based features are more important than contextual information based features, contextual information based features contribute very well in constructing the ranking model and help the translation relation based features for extracting translations more accurately.

In this research, we used different features for constructing a ranking model. In the future, we are going to define some other features and use them for improving the accuracy of constructed translation ranking model. We used LTR approach for improving the accuracy of translations extracted from different translation resources. Another interesting research direction could be using LTR approach for directly extracting translations from different corpora and using them for CLIR.

## Acknowledgment

This research is partially supported by ICT Research Institute (ITRC).

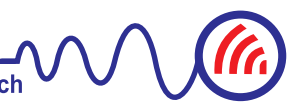

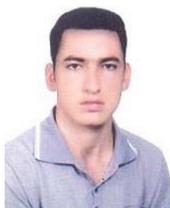

**Hosein Azarbonyad** is a researcher in the field of information retrieval. He finished his undergraduate studies in computer science at university of Tabriz and received his M.Sc. degree from university of Tehran. He is a member ofIntelligent Information Systems (IIS) Lab and Natural Language Processing (NLP) Lab in University of Tehran. His research interests include information retrieval, data mining, machine learning, natural language processing, and social networks.

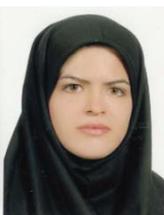

**Azadeh Shakery** is an assistant professor of Electrical and Computer Engineering at the College of Engineering, University of Tehran. She received her Ph.D in Computer Science from University of Illinois at Urbana-Champaign in 2008. Her research interests include text information management, information retrieval, and text and data mining.

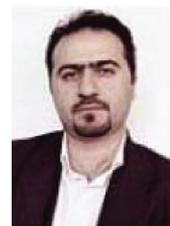

**Heshaam Faili** has his B.Sc. and M.Sc. in software engineering and Ph.D. in artificial intelligence from Sharif University of Technology. He is an assistant professor at Tehran University in the school of Electrical and Computer Engineering. His research interests include natural language processing, machine translation, data mining, and social networks.